\documentclass[aps,twocolumn,amsmath,amssymb,prl,showpacs,superscriptaddress]{revtex4-1}
\usepackage{amsfonts, hyperref, times}
\usepackage{graphicx}
\usepackage{color}

\begin{document}
  
\title{Spin texture motion in antiferromagnetic and ferromagnetic nanowires}

\author{Davi~R. Rodrigues}
\affiliation{
             Department of Physics \& Astronomy,
	    Texas A\&M University,
            College Station, Texas 77843-4242, USA
}
\author{Karin Everschor-Sitte}
\affiliation{Institute of Physics, Johannes Gutenberg Universit{\"a}t, 55128 Mainz, Germany}
\author{Oleg~A. Tretiakov}
\affiliation{Institute for Materials Research, Tohoku University, Sendai 980-8577, Japan}
\affiliation{School of Natural Sciences, Far Eastern Federal University, Vladivostok 690950, Russia}
\author{Jairo Sinova}
\affiliation{Institute of Physics, Johannes Gutenberg Universit{\"a}t, 55128 Mainz, Germany}
\affiliation{Institute of Physics ASCR, v.v.i, Cukrovarnicka 10, 162 00 Prag 6, Czech Republic}

\author{Ar.~Abanov}
\affiliation{
             Department of Physics \& Astronomy,
	    Texas A\&M University,
            College Station, Texas 77843-4242, USA
}

\date{\today}

\begin{abstract}
We propose a Hamiltonian dynamics formalism for the current and magnetic field driven dynamics of ferromagnetic and antiferromagnetic domain walls in one dimensional systems. To demonstrate the power of this formalism, we derive Hamilton equations of motion via Poisson brackets based on the Landau-Lifshitz-Gilbert phenomenology, and add dissipative dynamics via the evolution of the energy. We use this approach to study current induced domain wall motion and compute the drift velocity. For the antiferromagnetic case, we show that a nonzero magnetic moment is induced in the domain wall, 
which indicates that an additional application of a magnetic field would influence the antiferromagnetic domain-wall dynamics. We consider both cases of the magnetic field being parallel and transverse to the N{\'e}el field. Based on this formalism, we predict an orientation switch mechanism for antiferromagnetic domain walls which can be tested with the recently discovered N\'eel spin orbit torques.
\end{abstract}


\maketitle

\section{Introduction}

The insensitivity of antiferromagnets (AFM) to external magnetic fields, together with their inherent faster dynamics, affords these materials a technological advantage over their ferromagnetic (FM) counterparts \cite{Baryakhtar1979,Papanicolaou95}. However, the magnetic invisibility is responsible also for the difficulty in detecting and manipulating spin-textures that can store information in such materials. Making antiferromagnetic active components in spintronic devices is the focus of the emerging field of antiferromagnetic spintronics \cite{Jungwirth15,Gomonay2017}. 
%The  antiferromagnetic (AFM) spin textures have potential technological advantages over their ferromagnetic counterparts, as they are less susceptible to the external magnetic field. The studies of the AFM texture dynamics have been limited in the past [Junwwirth's review] \cite{Baryakhtar1979,Papanicolaou95} by the experimental difficulties to detect, control, and manipulate them.
This emerging field has seen a lot of recent progress in experimental techniques \cite{Tsoi2007,Jungwirth2011,Wu2011,Marti2012,Marti2014,Tsoi2014}, which has led to a raise in theoretical studies of the dynamics of AFM domain walls (DW) and other AFM spin textures   \cite{Brataas2011,Duine2011,Duine2011NatView,Tveten2013,Gomonay2010,Gomonay2013,Gomonay2016,Cheng2014,Kim2014,Cheng:2014gc,Cheng2015,Kim2015,Barker2015arXiv,Tveten2015,Tchernyshyov2017}.
Describing effectively the dynamics of these textures is a vital goal to connect experimental observables to the AFM textures and their dynamics. 

The analysis of FM DW dynamics in terms of a finite set of parameters have been considered recently in Refs. \cite{Tretiakov08,Clarke08}. The results obtained from such description based on a finite set of collective coordinates representing soft modes agree with experimental and numerical data. In this paper we show that, under some conditions, it is possible to combine such soft modes in terms of conjugated parameters and a Hamiltonian dynamics description. We apply this procedure to the electrical current and magnetic field driven dynamics of both FM and AFM DWs in nanowires. We derive the form of the effective Hamiltonian for DW dynamics based on the symmetries of the problem alone due to the transparent {\em Hamiltonian structure} of the effective equations of motion.

A strength of the Hamiltonian dynamics formalism provided in this paper is that it is insensitive to the details of the microscopic magnetic Hamiltonian.
%This formalism is very rich as it provides techniques to study several known aspects of AFM DW dynamics and introduces the analysis of new behaviors.
This aspect makes this formalism very powerful and rich as one can easily, with a few assumptions, study many known aspects of AFM DW dynamics and also include various interactions. Moreover, within the spirit of a phase space given by the conjugated soft modes, we may consider interactions and scattering of DWs, as well as thermodynamic effects.
As an application of the developed formalism, we will show that it is capable of describing: i) current-induced dynamics for both FM and AFM DWs, ii) magnetic field induced dynamics for both FM and AFM DWs, iii) orientation switching by current for AFM DWs. The orientation switching mechanism is a novel feature for AFM DWs which is naturally derived within our approach. The switch of configurations on antiferromagnets may have several applications to magnetic memory devices \cite{Wadley2016}. We also show below that other effects such as different anisotropies, nanowire inhomogeneity, etc., can be incorporated within the same formalism.

The paper is structured as follows. We will first describe the highly non linear dynamics of the magnetization field. We then define a simpler FM DW description through the use of proper canonical Hamiltonian variables. After deriving the Hamilton equations of motion for the spin polarized current driven FM DW in the dissipationless case, we show how the dissipation must be included in these equations. We also demonstrate how the form of the effective DW Hamiltonian can be understood from the symmetries of the problem. Finally, we apply this formalism to the case of AFM DWs, solve certain general cases for AFM DW dynamics, and make proposals for future AFM DW experiments. 

\section{Effective Hamiltonian description}

In the absence of spin-orbit torques (SOT) \cite{Hals2014,Khvalkovskiy2013,Kurebayashi2013,Chernyshov2008,Garello2013,Yu2014,Fan2014,Brataas2014,Kurebayashi2014,Ado2016} the magnetization dynamics of a FM due to magnetic fields and electric current is described by the Landau-Lifshitz-Gilbert (LLG) equation \cite{Tatara04, Thiaville05}:
\begin{equation}\label{eq:LLG}
\dot{\vec{m}}= \gamma_{0}\vec{m}\times \frac{\delta H_{m}}{\delta \vec{m}}-J\partial \vec{m}+\alpha \vec{m}\times \dot{\vec{m}}+\beta \vec{m}\times (J\partial \vec{m}) ,
\end{equation}
where $\vec{m}^{2}=1$ is the unitary vector along the magnetization, $\gamma_{0}$ is the gyromagnetic constant, the spatial derivative $\partial$ is along the nanowire, $H_{m}$ is a magnetic Hamiltonian, $\alpha$ is the Gilbert constant damping, and $\beta$ is a dimensionless damping parameter. $J$ corresponds to the spin-polarized current along the nanowire, with the amplitude given by
\begin{equation}
 J = \frac{jPg\mu_{B}}{2eM_{s}},
\end{equation}
where $j$ is the current density, $P$ is the polarization of the current, and $M_{s}$ is the saturation magnetization. The current, $\alpha $, and $\beta$ terms generally depend on the microscopic details of the full system. In particular, the dissipative $\alpha $ and $\beta $ terms depend on the strength and nature of the magnetization-electron and magnetization-phonon interactions. However, we assume locality (both in space and time) of the dissipative effects. The symmetry demands a specific form of the the dissipative terms and the current interaction. As the current, $\alpha$, and $\beta$ terms couple with the variation of the magnetization, at the scales of the magnetization configurations considered on this paper, only linear terms of these couplings are relevant. We also assume here the isotropic form of all current, $\alpha $ and $\beta $ terms, although this requirement can be violated in some materials. Nevertheless, generally the LLG equation has been shown to describe magnetization dynamics well in many different materials, see e.g. Ref.~\onlinecite{ThiavilleBook06}.

\begin{figure}
    \includegraphics[width=\linewidth]{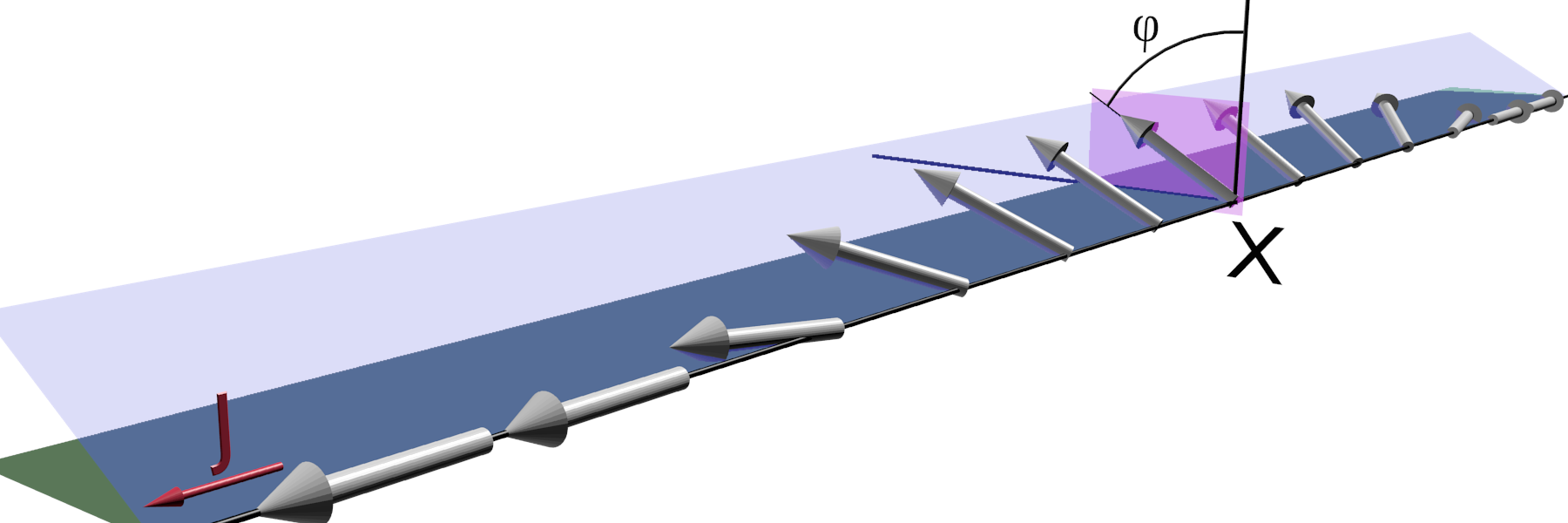}
    \caption{
        A sketch of ferromagnetic tail-to-tail domain wall. $X$ and $\phi $ denote the position and the tilt angle of the DW. $J$ corresponds to the injected spin polarized current. We assume the polarization of $J$ follows $
        \vec{m}$.
        \label{fig:FMDW}
    }
\end{figure}

The magnetic Hamiltonian $H_{m}$ includes all interactions for the local magnetization in the ferromagnet, such as the exchange interaction and all the magnetic anisotropies. It may also include Dzyaloshinskii-Moriya, dipole-dipole, and other interaction terms. We assume that $H_{m}$ does not explicitly depend on time.

\subsection{Domain wall motion in a ferromagnet}

A typical transverse DW in a ferromagnetic nanowire is well approximated by a rigid object, which can move along and rotate around the wire axis \cite{Tretiakov_DMI}. During this motion the DW shape changes a little, since in the presence of anisotropy the modes corresponding to the change of the DW shape are gapped. Here we derive the dynamics of the DW based on the Hamiltonian equations of motion. To do so we calculate the Poisson brackets for two parameters of this object.

The Poisson brackets of the unit vector $\vec{m}$, a representation of SO(3), are
\begin{equation}
 \{m_{i}(x), m_{j}(x') \}=\epsilon_{ijk}m_{k}(x)\delta (x-x').
\end{equation}

The magnetization configuration $\vec{m}(x)$ of a single DW in a nanowire may be described in terms of the soft modes $X$, the position of the domain wall, and $\phi$, the rotation of magnetization at the DW center around the nanowire axis, see Fig.~\ref{fig:FMDW}. A variation of the configuration in terms of these parameters corresponds to
\begin{equation}
 dm_{i}(x)=-dX \partial m_{i}(x)+d\phi \epsilon_{ijk}e_{j}m_{k}(x),
\end{equation}
where $\vec{e}$ is the unit vector along the wire. Comparing the total volume of the phase space in terms of $\vec{m}$ and variables $X$ and $\phi $, we find 
\begin{equation}\label{eq:Poisson}
\{X,\phi  \}=\pm 1,  
\end{equation}
where the $+$ and $-$ signs are for the tail-to-tail and head-to-head DWs, respectively \footnote{We set the lattice constant to $1$ throughout this paper.}. Note that for a FM DW, the DW position $X$ and the angle $\phi$ are canonically conjugated Hamiltonian variables independent of the microscopic form of the Hamiltonian.

Let us now consider the dependence of the total magnetic energy on time. As the Hamiltonian does not depend explicitly on time, we can write 
\begin{eqnarray} 
&&\dot{E}=\int \frac{\delta H_{m}}{\delta m_{i}(x)}\dot{m}_{i}dx  
\end{eqnarray}
and substitute Eq.~\eqref{eq:LLG} for $\dot{\vec{m}}$. Upon doing this we notice that only one of the terms is not dissipative, $-J\int \frac{\delta H_{m}}{\delta m_{i}(x)}\partial m_{i}dx = J\partial_{X}E$, where $X$ is the DW center coordinate. The remaining terms are first order in dissipation. Using $\epsilon_{ijk}m_{j} \frac{\delta H_{m}}{\delta m_{k}}=\dot{m}_{i}+J\partial m_{i}$ we find
\begin{equation}\label{eq:Edot}
\dot{E}-J\partial_{X}E= - \int (\dot{m}_{i}+J\partial m_{i}) (\alpha\dot{m}_{i}+ \beta J\partial m_{i} ) dx .
\end{equation}

First we consider the dissipationless dynamics, where we set $\alpha =\beta =0$. We obtain then the equation of Hamiltonian dynamics $\dot{E}=J\partial_{X}E$. Therefore, there exists an effective Hamiltonian $H(X,\phi )$ such that 
$ \dot{E}=\{E,H \}=J\partial_{X}E $.
In the absence of current, the Hamiltonian must be equal to the magnetic energy $E(X,\phi)$ of the DW, and thus we conclude
\begin{equation}\label{eq:Ham}
H(X,\phi )=E(X,\phi )\pm J\phi  ,
\end{equation}
as this Hamiltonian gives the correct $\dot{E}$. This energy function is all one needs to know to obtain the effective equations of motion for the FM DW.

The Hamiltonian equations for the coordinates $X$ and $\phi $ take the form
\begin{equation}
 \dot{X}=\{X,H \}=\pm \frac{\partial H}{\partial \phi },\qquad  \dot{\phi }= \{\phi ,H \}=\mp \frac{\partial H}{\partial X}.
\end{equation}
Next, we include dissipative terms $\gamma_{X}$ and $\gamma_{\phi}$ in these equations:
\begin{equation}\label{eq:Hamdiss}
 \dot{X}=\{X,H \}+\gamma_{X},\qquad   \dot{\phi }= \{\phi ,H \}+\gamma_{\phi }.
\end{equation}
To calculate the dissipative terms, we expand Eq.~\eqref{eq:Edot} for the DW motion to the linear order in $J$
\begin{eqnarray}
 \dot{E}-J\partial_{X}E &=& -\alpha \dot{X}^{2}\Delta_{X}^{-1}
+2\alpha\dot{\phi }\dot{X} \Gamma 
-\alpha\dot{\phi }^{2} \Delta_{\phi }\nonumber\\
&&
+(\alpha+\beta ) J \dot{X}\Delta_{X}^{-1}
-(\alpha+\beta )\dot{\phi } J \Gamma ,
\label{eq:diss}
\end{eqnarray}
where the constants are defined as
\begin{eqnarray} 
&&\Delta_{X}^{-1}= \int (\partial \vec{m})^{2}dx,\quad \Delta_{\phi }=\int (1-m^{2}_{x})dx, 
\nonumber\\
&&
{\rm and}\,\,\Gamma =\int [\vec{e}\times \vec{m}] \partial\vec{m}dx.\nonumber
\end{eqnarray} 
They are the parameters determining the DW dynamics and depend on the DW shape. For a planar DW, when the magnetic Hamiltonian has no Dzyaloshinskii-Moriya term \cite{Tretiakov_DMI}, the constant $\Gamma$ vanishes, $\Gamma = 0$.

We compare now Eq.~\eqref{eq:diss} to the same expression we computed using \eqref{eq:Hamdiss},
 $ \dot {E}- J\partial_{X}E=\mp J\gamma_{\phi } \pm \gamma_{\phi }\dot{X}\mp \gamma_{X}\dot{\phi } $,
and find for the dissipative terms
\begin{eqnarray}
\gamma_{\phi }&=&  \mp \alpha \dot{X}\Delta_{X}^{-1}\pm \beta J \Delta_{X}^{-1},\\
\gamma_{X }&=&  \pm 
\alpha\dot{\phi } \Delta_{\phi } \mp 2\alpha\dot{X}\Gamma \pm (\alpha + \beta) J \Gamma.  
\end{eqnarray}
We note that Eqs.~\eqref{eq:Hamdiss}, together with the constants $\gamma_{X}$ and $\gamma_{\phi }$ defined above, completely determine the DW dynamics in a ferromagnetic nanowire in terms of $\alpha $, $\beta $, and the parameters of the magnetic Hamiltonian. In particular, many previous results on ferromagnetic DW dynamics \cite{Tretiakov08, Tretiakov_DMI, Tretiakov:losses, Liu11, Tretiakov2012} can be easily obtained within this formalism.

We emphasize that although in this paper we assume the direction of the uniaxial magnetic anisotropy to be along the nanowire, this description is valid for any other anisotropy direction. The angle $\phi $ is then just the angle of the central spin in the DW, which rotates around the anisotropy axis.  In the case of anisotropy axis perpendicular to the nanowire this angle describes the oscillations between the Bloch and N{\'e}el DWs.

\begin{figure}
    \includegraphics[width=\linewidth]{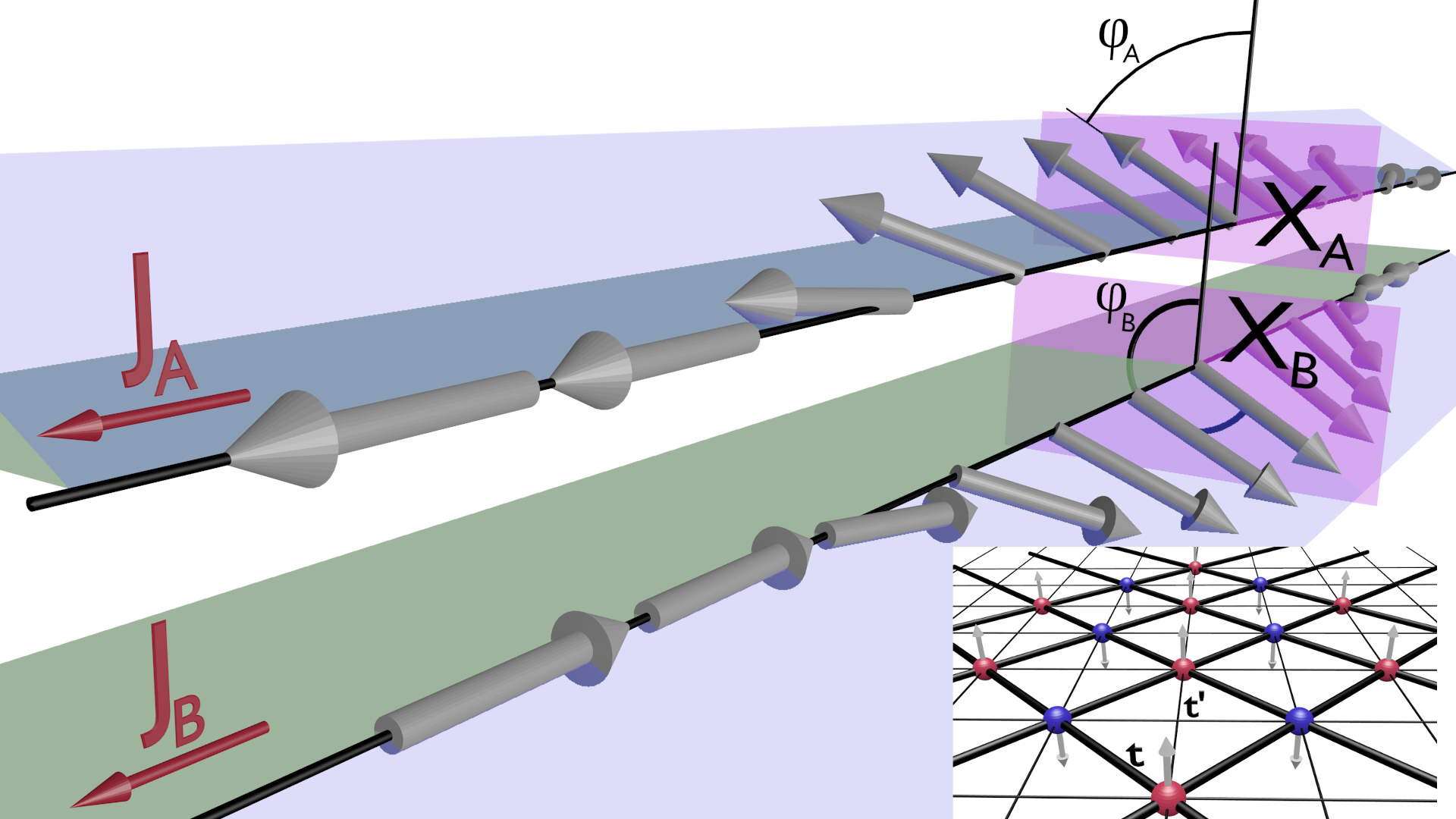}
    \caption{
        A sketch of antiferromagnetic domain wall. The AFM DW is a composite object consisting of two ferromagnetic DWs on two sublattices ($A$ and $B$) with respective position $X_A$ ($X_B$) and the tilt angle of the ferromagnetic DW $\phi_A$ ($\phi_B$) on respective sublattice. One ferromagnetic DW is tail-to-tail ($A$) and the other is head-to-head ($B$). 
%The currents $J_A$ and $J_B$ on the sublattices $A$ and $B$ can be different. 
The inset shows the AFM lattice with up- and down-spin sublattices and nearest (second nearest) neighbor hopping constants $t$ ($t^{\prime}$).
        \label{fig:AFMDW}
    }
\end{figure}

\subsection{Domain wall motion in an antiferromagnet}

Below we show how to extend the Hamiltonian formalism of DW dynamics to the important case of AFM DW. Both theoretical and experimental studies \cite{Brataas2011, Duine2011, Duine2011NatView, Tveten2013, Cheng2014, Kim2014, Cheng:2014gc, Cheng2015, Kim2015, Barker2015arXiv, Tveten2015, Tsoi2007, Jungwirth2011, Wu2011, Marti2012, Marti2014, Tsoi2014} of the AFM dynamics have been very active recently due to improved experimental capabilities to detect, produce, and manipulate these systems. However, a more systematic approach to the motion of AFM DWs is still missing and here we make an attempt to fill in this gap. 

We consider a collinear AFM on a bipartite lattice with sublattices $A$ and $B$. A typical nanowire is quasi one-dimensional from the point of view of magnetization and magnetization dynamics, but still is three-dimensional from the point of view of the electronic degrees of freedom. An electron has hopping matrix elements both between the nearest neighbors and between next nearest neighbors on the lattice. In the presence of the AFM order in the case of large on-site electron spin-magnetization interaction the hopping between the nearest neighbors -- between a site of sublattice $A$ and a nearest site in sublattice $B$ -- is suppressed, while the hoping between the next nearest neighbors -- the sites of the same sublattice -- is unaffected. Thus, we can assume that each electron lives on its own sublattice and interacts only with the magnetization of the same sublattice. A hybrid systems of bilayer materials, i.e. what is known as an artificial antiferromagnet, where both layers have ferromagnetic in-layer order coupled antiferromagnetically to each other is a good example of such a description.

In the presence of an AFM DW this separation of the electrons to different sublattices is not perfect, but under the assumption that the DW width is much larger than the electronic coherence length, the corrections to this picture are expected to be small.  
It is important to note that, if there is a percolation within each sublattice, the potential difference across the wire is the same on both sublattices, however the resistivity of each of them does not have to be the same. Thus the electrical currents on the two sublattices can be different. This case is especially crucial in the situation where the sublattices are made of different atoms, as it is on some hybrid systems. To simplify the equations, however, we assume that, given the assumptions of the scales considered, the currents on the two sublattices are the same \footnote{The case of $J_{A}\not=J_{B}$ will be considered in \cite{Davi_thesis}.}.

The above described situation leads to the following picture. A single AFM DW is composed of two coupled FM DWs on each of the sublattices with its own current. One of the DWs is tail-to-tail (DW on sublattice A in Fig.~\ref{fig:AFMDW}) and the other one is head-to-head. We thus introduce two sets of variables, $X_{A}$ and $\phi_{A}$ for the head-to-head DW on sublattice $A$ and $X_{B}$ and $\phi_{B}$ for the tail-to-tail DW on sublattice $B$. To simplify the expressions below we assume that the DW parameters $\Delta_{\phi }$, $\Delta_{X}$, $\alpha $, and $\beta $ are the same for the two DWs, and that the DWs are planar, i.e. $\Gamma =0$. 

The magnetic Hamiltonian depends now on the two sets of coordinates $H(X_{A},\phi_{A},X_{B},\phi_{B})$. 
The center of the DW is at $X=\frac{1}{2}(X_{A}+X_{B})$. The value of $X_{A}-X_{B}$ gives the magnetization of the AFM DW along the nanowire. $\phi_{B}-\phi_{A}$ is the angle between the directions of the spins at the center of the DWs. It corresponds to the magnetic moment of the AFM DW perpendicular to the AFM DW plane. The coupled equations of motion are:
\begin{eqnarray} 
\dot{X}_{A,B}&=&\{X_{A,B},H \} \mp \alpha\dot{\phi }_{A,B} \Delta_{\phi },\\  
%\dot{X}_{A}&=&\{X_{A},H \} -\alpha\dot{\phi }_{A} \Delta_{\phi },\\  
%\dot{X}_{B}&=&\{X_{B},H \} +\alpha\dot{\phi }_{B} \Delta_{\phi },\\
\dot{\phi }_{A,B}&=&\{\phi_{A,B} ,H \}\pm  \frac{\alpha}{\Delta_{X}}\dot{X}_{A,B}\mp  \frac{\beta}{\Delta_{X}}J_{A,B},
%\dot{\phi }_{A}&=&\{\phi_{A} ,H \}+ \frac{\alpha}{\Delta_{X}}\dot{X}_{A}- \frac{\beta}{\Delta_{X}}J_{A},\\ 
%\dot{\phi }_{B}&=&\{\phi_{B} ,H \}- \frac{\alpha}{\Delta_{X}}\dot{X}_{B}+ \frac{\beta}{\Delta_{X}}J_{B}
\end{eqnarray}
where upper and lower signs are for sublattices $A$ and $B$, respectively, and the Poisson brackets are  
 $\{X_{A},\phi_{A}  \}= -1$, $\{X_{B},\phi_{B}  \}= 1$. The Hamiltonian $H$ is
\begin{eqnarray} 
\label{Hamilt}
H(X_{A},\phi_{A},X_{B},\phi_{B} )\!&=&\! E(X_{A},\phi_{A},X_{B},\phi_{B} ) -hX_{A}+hX_{B}\nonumber\\
&&- J_{A}\phi_{A}+J_{B}\phi_{B},
\end{eqnarray}
where we have explicitly written the coupling of the DWs on the sublattices to the magnetic field $h$, and $E$ is then the energy of the two DWs in the absence of the magnetic field. It includes the magnetic interaction of the two DWs and thus depends on the two sets of coordinates. In a translationally invariant nanowire the energy $E$ is independent of $X_{A}+X_{B}$. However, in the problems for the DW dynamics in a nanowire with a non-uniform shape \cite{Tretiakov2012}, one should add a term 
$\Omega(X_{A}^2+X_{B}^2)/2$ to the energy $E$, where $\Omega$ is a constant inversely proportional to the nanowire curvature. 

One can write the function $E(X_{A},\phi_{A},X_{B},\phi_{B}) $ in a very general form. We notice that the minimum of the total AFM DW energy is reached at $X_{A}-X_{B}=0$ and $\phi_{A}-\phi_{B}=\pi $. Expanding for small $|X_{A}-X_{B}|$ and the first harmonic of the angular dependence, we find
\begin{equation}\label{eq:energy}
E(X_{A},\phi_{A},X_{B},\phi_{B})=\frac{\Delta_{1}}{2}(X_{A}-X_{B})^{2}+\Delta_{2}\cos (\phi_{A}-\phi_{B}).  
\end{equation}
The constants $\Delta_{1}$ and $\Delta_{2}$ are of the order of $J_{AF}\Delta_{X}^{-2}$ and $J_{AF}$, respectively, where $J_{AF}$ is the antiferromagnetic exchange constant. One also can add a transverse anisotropy by adding $K(\sin^{2}\phi_{A}+\sin^{2}\phi_{B})$ to the energy.

 Using the Hamiltonian (\ref{Hamilt}) with energy \eqref{eq:energy} to calculate the Poisson brackets, we obtain 
%\begin{eqnarray}
%\label{X1}
%\dot{X}_{A}&=& \Delta_{2}\sin (\phi_{A}-\phi_{B})-\alpha\dot{\phi }_{A}\Delta_{\phi } +J, \\
%\dot{\phi }_{A}&=& \Delta_{1}(X_{A}-X_{B})-h+\frac{\alpha}{\Delta_{X}}\dot{X}_{A}-\frac{\beta}{\Delta_{X}}J, \\
%\dot{X}_{B}&=& \Delta_{2}\sin (\phi_{A}-\phi_{B})+\alpha\dot{\phi }_{B}\Delta_{\phi } +J, \\ 
%\dot{\phi }_{B}&=& \Delta_{1}(X_{A}-X_{B})-h-\frac{\alpha}{\Delta_{X}}\dot{X}_{B}+\frac{\beta}{\Delta_{X}}J.
%\label{phi2}
%\end{eqnarray}
\begin{eqnarray}
\label{X1}
\dot{X}_{A,B}&=& \Delta_{2}\sin (\phi_{A}-\phi_{B})\mp \alpha\dot{\phi }_{A,B}\Delta_{\phi } +J, \\
\dot{\phi }_{A,B}&=& \Delta_{1}(X_{A}-X_{B})-h\pm \frac{\alpha}{\Delta_{X}}\dot{X}_{A,B}\mp \frac{\beta}{\Delta_{X}}J, 
\label{phi2}
\end{eqnarray}
where we set $J_{A}=J_{B}=J$. We point out that for AFM materials with different compositions of their sublattices $J_{A}\not=J_{B}$. In the limiting case that one sublattice is insulating, passing the current through this material would lead to the rotation of the AFM DW as in the case of a ferromagnetic DW above the Walker breakdown. In this paper we assume that the easy-axis magnetic anisotropy is along the wire, and the magnetic field $h$ in the above equations is also along the wire -- along the anisotropy axis. However, the general formalism is valid for any direction of the anisotropy after a trivial change of notations.

The equations of motion \eqref{X1} and \eqref{phi2} are one of our main results. They provide the dynamics of AFM DW interacting with both magnetic field $h$ and electrical current $J$. These two interactions may be studied independently.

{\it Current induced AFM DW dynamics.}- Many aspects of the current driven dynamics of magnetization configurations in antiferromagnetic is well known, see for example Refs. \cite{Gomonay2013,Gomonay2010,Barker2015arXiv,Wadley2016,Tchernyshyov2017}. We show that some results may be easily derived within the Hamiltonian formalism.
From Eqs.~\eqref{X1} -- \eqref{phi2} we consider next  the case with no magnetic field, i.e. $h=0$, and
\begin{eqnarray} 
\label{X12}
&& \dot{X}_{A}- \dot{X}_{B}=-\alpha (\dot{\phi}_{A}+\dot{\phi }_{B})\Delta_{\phi },\\
&& \dot{\phi}_{A}+\dot{\phi }_{A}=2\Delta_{1}(X_{A}-X_{B})+\frac{\alpha}{\Delta_{X}}(\dot{X}_{A}-\dot{X}_{B}).
\label{phiX}
\end{eqnarray}
As $\dot{X}_{A}- \dot{X}_{B}$ is already first order in the dissipation, the second term on the right-hand side (RHS) of Eq.~(\ref{phiX}) should be dropped.  Then, by substituting Eq.~(\ref{phiX}) into (\ref{X12}) we obtain, 
\begin{equation}
 \dot{X}_{A}- \dot{X}_{B}= -2\alpha \Delta_{\phi } \Delta_{1}(X_{A}-X_{B}).
 \end{equation}
It has an exponentially decaying solution for $X_{A}-X_{B} \propto \exp ( -2\alpha \Delta_{\phi } \Delta_{1} t)$ and thus in the steady state $\dot{\phi}_{A}=-\dot{\phi}_{B}$. 
We also obtain the other two equations from the system (\ref{X1}) - (\ref{phi2}):
\begin{eqnarray} 
&& \dot{\phi }_{A}- \dot{\phi }_{B}=\frac{2\alpha }{\Delta_{X}}\dot{X}-\frac{2\beta }{\Delta_{X}}J,
\nonumber\\
&&\dot{X}= \Delta_{2}\sin (\phi_{A}-\phi_{B})-\alpha \Delta_{\phi }\frac{\dot{\phi }_{A}-\dot{\phi }_{B}}{2}+J .\nonumber
\end{eqnarray}
Similarly $\dot{\phi}_{A}-\dot{\phi}_{B}$ is already of the first order in dissipation, so we can neglect the second term on the RHS of the second equation to find:
\begin{eqnarray} 
&&\dot{X}= \Delta_{2}\sin (\phi_{A}-\phi_{B})+J.
\end{eqnarray}

This set of equations has the simple solutions $\dot{\phi }_{A}- \dot{\phi }_{B}=0$ and $\dot{X}=V$ given by
\[ 
V=\frac{\beta}{\alpha }J,\qquad \sin (\phi_{A}-\phi_{B})= -\frac{J}{\Delta_{2}}(1-\beta /\alpha ).  
\]
There is a critical current $J_{c}=\frac{\alpha \Delta_{2}}{|\alpha -\beta |}$ up to which this solution exists. This current is generally large, since $\Delta_{2}$ is of the order of the exchange constant. Physically this critical current corresponds to the situation when the magnetizations on the two sublattices rotate with respect to each other and eventually point in the same direction.  For small currents $J\ll J_{c}$ this solution shows that a moving AFM DW is not rotating and has a magnetic moment of the order of $\Delta_{X}\frac{J}{J_{c}}$ perpendicular to the plane of the AFM DW. 

For $J>J_{c}$ the AFM DW will not rotate, but the magnetic moment perpendicular to the AFM DM oscillates in time with $T=\frac{\Delta_{X}}{2\alpha \Delta_{2}}\frac{2\pi }{\sqrt{(J/J_{c})^{2}-1}}$. The AFM DW velocity is also not constant in time. The average AFM DW velocity is given by $\langle V\rangle = J-\frac{J\Delta_{2}}{J_{c}}\left(1-\sqrt{(J/J_{c})^{2}-1} \right)$.

{\it Magnetic field parallel to the nanowire.}- Next, we consider the AFM DW dynamics under the action of the magnetic field along the wire in the absence of current. In this situation we find from Eqs.~(\ref{X1}) -- (\ref{phi2}) that in the steady state $\dot{\phi}_{A}=\dot{\phi}_{B}=0$, and $\phi_{B}=\pi +\phi_{A}$ (a small deviation from this decays exponentially with time), while $\dot{X}_{A}=\dot{X}_{B}=0$, and $X_{A}-X_{B}=h/\Delta_{1}$, which show that the induced magnetic moment is $\sim h\Delta_{X}/\Delta_{1}$. Also, notice that unlike in a FM DW, there is no motion induced by an external magnetic field in this configuration.

{\it Magnetic field perpendicular to the nanowire with current.}- Let us consider the AFM DW dynamics under the magnetic field perpendicular to the current, i.e. the nanowire axis. This magnetic field couples to the angles $\phi_{A}$ and $\phi_{B}$, the corresponding term in the Hamiltonian is $-h(\sin \phi_{A}+\sin \phi_{B})$, where $h$ is the magnetic field multiplied by the perpendicular magnetization of a single DW. Such term in the Hamiltonian does not change the equations for $\dot{\phi}_{A}$ and $\dot{\phi}_{B}$, see Eqs.~(\ref{X1}) -- (\ref{phi2}), but adds a term $h\cos \phi_{A}$ to the RHS of equation for $\dot{X}_{A}$ and $-h\cos \phi_{B}$ to the RHS of equation for $\dot{X}_{B}$. Then, it follows,
\begin{eqnarray}
\label{X2}
\dot{X}_{A,B} &=& \Delta_{2}\sin (\phi_{A}-\phi_{B})\mp \alpha\dot{\phi }_{A,B}\Delta_{\phi } + J \pm h\cos \phi_{A,B}\nonumber\\
\\
\dot{\phi }_{A,B}&=& \Delta_{1}(X_{A}-X_{B}) \pm \frac{\alpha}{\Delta_{X}}\dot{X}_{A,B}\mp \frac{\beta}{\Delta_{X}}J\,\,.
\label{phi3}
\end{eqnarray}
A simple solution for the steady state at small $J$ and $h$ is given by $X_{A}=X_{B}$ and  $\dot{X}_{A}=\frac{\beta}{\alpha}J$ for the coordinates. Note that this is the same result as for the current driven AFM DW motion. For the angles, we obtain $\phi_{A} = \pi + \phi_{B}$ and
\begin{equation}\label{eq:phiA}
 \cos \phi_{A} = -\frac{J}{h}(1-\beta /\alpha).
\end{equation}

We notice from Eqs.~(\ref{X2}) -- (\ref{phi3}) that as we switch the current with constant magnetic field, the angles could also be switched. 

{\it AFM DW orientation switching mechanism.}- Given a static AFM DW without current or magnetic fields applied to it, from Eqs.~(\ref{X1}) -- (\ref{phi2}) the configuration is given by $X_{A}=X_{B}$ and $\phi_{A} = \phi_{B}$ or $\phi_{A} = \phi_{B} + \pi$. The first corresponds to a ferromagnetic state. The second configuration, as we apply a magnetic field perpendicular to the nanowire, may represent two different drift velocities, see Eqs.~(\ref{X1}) -- (\ref{phi2}). These different behaviors suggest that it would be interesting to study a reorientation mechanism for an AFM DW.

First, we notice that a weak magnetic field applied parallel to the nanowire would induce a precession of the AFM DW. This precession, however, would be extremely slow. As we are interested in practical use for the reorientation, we need to consider faster processes. This may be obtained by considering a time dependent current and a magnetic field perpendicular to the nanowire.

\begin{figure}
    \includegraphics[width=\linewidth]{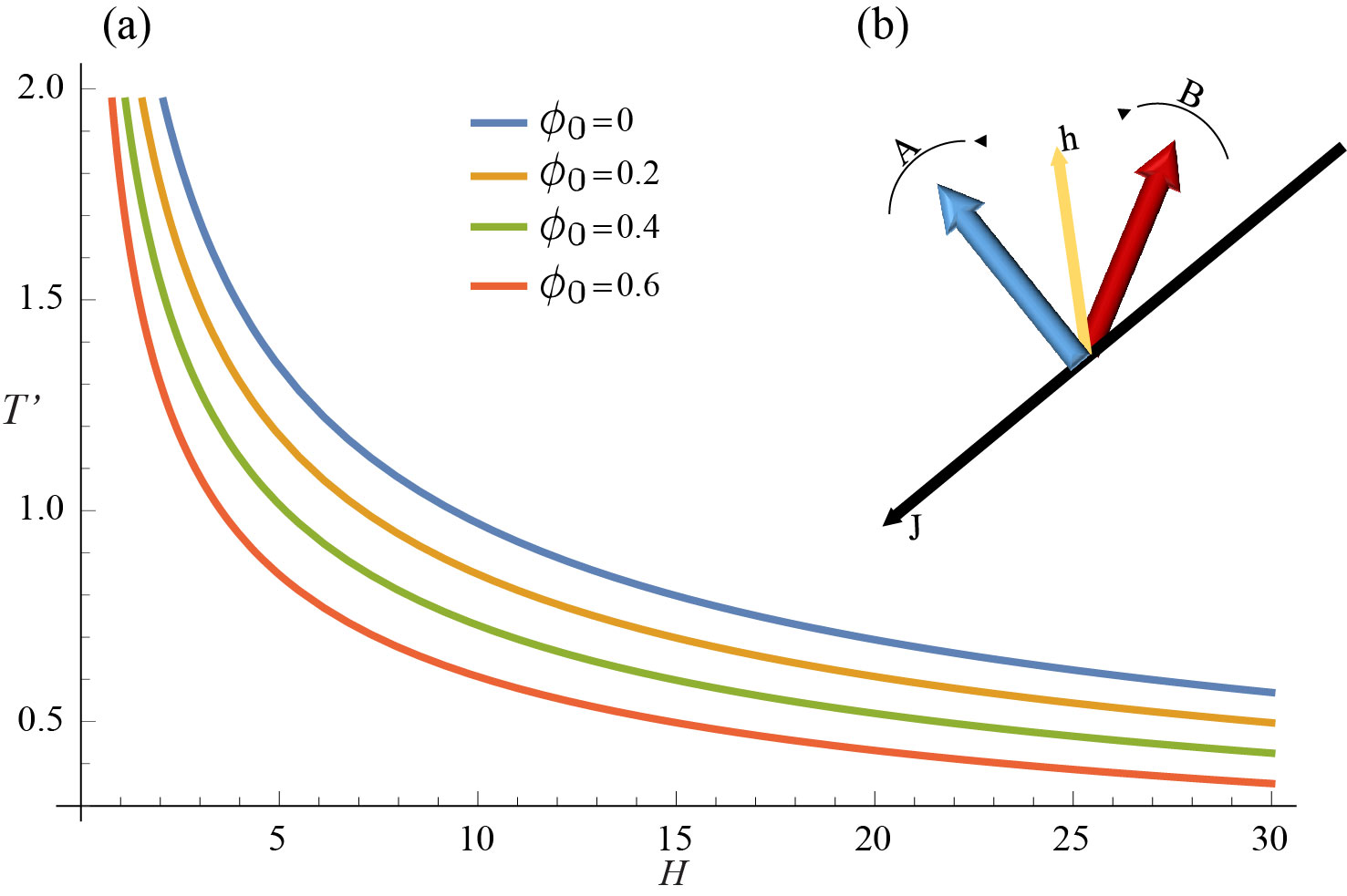}
    \caption{
        (a) A plot of the numerical solution of integral \eqref{periodT} giving the time of switching $T' = T\gamma/\Delta_{X}$ as a function of the parameter $H = E_{T}/\gamma^2$ for different initial angles $\phi_{0}$. As expected, the time of switching decreases as we increase the initial $\dot{\phi}$. (b) A sketch of the reorientation mechanism for AFM DWs. The different sublattices' orientation will cross, producing a temporary magnetic moment perpendicular to the wire during the process.
        \label{fig:ET}
    }
\end{figure}

To illustrate the reorientation mechanism, we consider that initially $X_{A} = X_{B} = X$ and $\phi_{A} = \pi - \phi_{B} = \phi$. From Eqs.~(\ref{X2}) --  (\ref{phi3}), we note that these relations are valid for the entire switching process as long as we apply a time-dependent current of the form
\begin{eqnarray}\label{jphi}
J(t) = \frac{1}{\alpha-\beta}\Big[\Delta_{X}\dot{\phi}+\alpha\cos\phi\left(2\Delta_{2}\sin\phi-h\right)\Big],
\end{eqnarray}
where we neglected the terms proportional to $\alpha^2$. Other types of currents may be considered for the reorientation. However, the dynamics involved will be a lot more complex and may not be possible to obtain the exact time dependence analytically. For initial angle $\phi_0$ we consider that we have a static AFM DW profile with no current. This implies
\begin{equation}\label{phi0}
 \sin\phi_{0} = \frac{h}{2\Delta_{2}}.
\end{equation}
In order to obtain the minimum Ohmic losses for the switching process with finite time of switching $T$, we find that the time dependence of $\phi$ is given by
\begin{equation}\label{phit}
t = \frac{\Delta_{X}}{\gamma}\int_{\phi_{0}}^{\phi(t)} \frac{d\phi}{\sqrt{\frac{E_{T}}{\gamma^2} + \cos^2\phi\left(\sin\phi-\sin\phi_{0}\right)^2}},
\end{equation}
where $\gamma = 2\Delta_{2}\alpha$ and $E_{T} = \Delta_{X}^2\dot{\phi}_{0}^2$ is the constant related to the time of switching $T$ by
\begin{equation}\label{periodT}
T = \frac{\Delta_{X}}{\gamma}\int_{\phi_{0}}^{\pi-\phi_{0}} \frac{d\phi}{\sqrt{\frac{E_{T}}{\gamma^2} + \cos^2\phi\left(\sin\phi-\sin\phi_{0}\right)^2}}.
\end{equation}
Therefore, with the current given by Eq.~(\ref{jphi}) and Eq.~(\ref{phit}) satisfying Eq.~(\ref{phi0}), we are able to efficiently switch the orientation of the AFM DW within a finite time  $T$ given by Eq.~(\ref{periodT}). It is important to notice, however, that with the time dependent current, Eq.~(\ref{jphi}), it is also possible to obtain other $\phi(t)$ evolutions that also corresponds to the switching process, but with higher energy losses. The integral can be solved in terms of elliptic functions, see Fig.~\ref{fig:ET}(a). Once we are able to switch the  orientation in a controllable fashion, one must then measure the switching of this 180 AFM DW. Whereas most measurements cannot directly observe such DW, a system that contains the newly discovered Neel spin-orbit torque, \cite{Gomonay2016,Wadley2016}, can in principle be sensitive to the orientation of a 180 DW as shown in Ref.~\cite{Gomonay2016}. The measurement of the DW orientation was also considered in Refs.~\cite{Vedmedenko2004,Liu11}. One must also notice that during the switching process a finite magnetic moment arises, see Fig.~\ref{fig:ET}(b), which can be measured experimentally allowing an indirect measure of the process.

In the present formulation the direction of the magnetic field in the plane perpendicular to the nanowire is arbitrary. However, in the presence of the  transverse anisotropy this effect will be the largest if the field direction is perpendicular to the transverse anisotropy axis. The current required for the switching in this case will be determined by both the magnitude of the magnetic field and by the anisotropy.

\section{Conclusion}

We have developed a Hamiltonian approach to the current and magnetic field driven dynamics of both ferromagnetic and antiferromagnetic DWs, which describes the domain walls as rigid topological objects. We have shown how dissipation is included in this description by means of LLG equation formalism. The dynamics can be described by a set of universal equations which depend only on a few parameters. These parameters can be measured in real nanowires either through magnetoresistance or through electrical means as shown in Ref. [\onlinecite{Liu11}].

We have shown that the developed formalism allows to solve various problems of both FM and AFM DW dynamics on the same footing and extend it to different geometries. As the Hamiltonian formalism does not depend on microscopic aspects, it allows to easily introduce new interactions. In particular, it can be used to describe both FM and AFM DW dynamics induced by parallel magnetic field and current. As a consequence of this analysis, we were able to obtain a novel orientation switch mechanism for AFM DWs. With the developments of measuring techniques, the mechanism described here may be useful for memory devices.

\section{Acknowledgments}

Ar. A. is very grateful to the INSPIRE group in Johannes Gutenberg-Universit{\"a}t, Mainz, Germany. We are thankful to H.~Gomonay for valuable discussions, to O.~Tchernyshyov for explaining to us the Poisson brackets \eqref{eq:Poisson} as the Poisson brackets of the component of the total angular momentum and the corresponding angle and to B.~McKeever for reviewing the calculations on this paper. O.A.T. acknowledges support by the Grants-in-Aid for Scientific Research (Nos. 25247056, and 15H01009) from the Ministry of Education, Culture, Sports, Science and Technology (MEXT) of Japan. K. E.-S. acknowledges funding from the German Research Foundation (DFG) under the Project No. EV 196/2-1. J. S. acknowledges funding from the Alexander von Humboldt Foundation, the ERC Synergy Grant SC2 (No. 610115), the Transregional Collaborative Research Center (SFB/TRR) 173 SPIN+X, and Grant Agency of the Czech Republic grant no. 14-37427G.

%\bibliography{micromagnetics}
\bibliography{afm_references}

\end{document}